\documentstyle[aps,epsf]{revtex}

\newcommand{\ket}[1]{\mathop{\left| #1 \right\rangle}\nolimits}

\begin{document}

\title{Generation of entanglement in a system of two dipole-interacting atoms
by means of laser pulses}
\author{I.V. Bargatin, B.A. Grishanin, and V.N. Zadkov}
\address{International Laser Center and Department of Physics,\\
M.\ V.\ Lomonosov Moscow State University,  Moscow 119899, Russia} \maketitle

\begin{abstract}
Effectiveness of using laser field to produce entanglement between two
dipole-interacting identical two-level atoms is considered in detail. The
entanglement is achieved by driving the system with a carefully designed laser
pulse transferring the system's population to one of the maximally entangled
Dicke states in a way analogous to population inversion by a resonant
$\pi$-pulse in a two-level atom. It is shown that for the optimally chosen
pulse frequency, power and duration, the fidelity of generating a maximally
entangled state approaches unity as the distance between the atoms goes to
zero.
\end{abstract}

\bigskip

With recent experimental advances in the methods for coherent manipulation of
quantum system on the level of individual particles, many of previously purely
speculative problems become surprisingly up-to-date. In particular, much
activity of physicists from different research fields is currently devoted to
clarification of the entanglement concept \cite{preskill}, ways for its
quantification, purification, and creation. Casual creation of entangled states
of atoms by coherent manipulation with light currently poses one of the biggest
challenges in the field of quantum optics  \cite{atoment}. Conversely, the
resonant dipole-dipole interaction (RDDI) and cooperative relaxation effects
associated with it are rather traditional topics of research. Recently the RDDI
has been investigated as a source of interference phenomena in emission spectra
\cite{agarwal81,meystre93,kurizki87}, super- and sub-radiance
\cite{brewer95,andreevbook}, photon bunching \cite{beige98}, collisions in the
laser cooling processes \cite{smith91,meystre96}, and as a mechanism for
realizing quantum logical gates \cite{barenco95,brennen98}). In this paper we
address the question: how effectively can the RDDI along with coherent laser
pulses be used for creation of multi-atomic entangled states.

\medskip
In our model two identical two-level atoms are located at a fixed distance $R$
and can be driven by laser beam that is either parallel or perpendicular
(according to geometries identified, respectfully, as antisymmetric and
symmetric, Fig. 1) to the radius vector $\vec R$ connecting the atoms. Within
the interaction picture and rotating wave approximation, the evolution is
described by the following master equation \cite{kurizki87}:
\begin{equation}\label{ME}
  \frac{\partial \hat \rho}{\partial t}=
  -\frac{i}{h}\left[\hat{\cal H}_{\rm eff},\rho\right]+
  \sum_{i,j=1,2}\frac{\gamma_{ij}}{2}(2\hat \sigma^-_i\hat\rho\hat \sigma^+_j -
  \hat\rho\hat \sigma^-_i\hat
  \sigma^+_j - \hat \sigma^-_i\hat \sigma^+_j\hat\rho ),
\end{equation}
where
\begin{equation}\label{Heff}
\hat{\cal H}_{\rm eff}=\frac{\hbar}{2} \left[-\delta \left(\hat \sigma^z_1+\hat
\sigma^z_2\right)+ \Omega_1\hat \sigma^+_1+\Omega_2\hat \sigma^+_2+ \chi\hat
\sigma^+_1\hat \sigma^-_2+ \mbox{h.\,c.}\right].
\end{equation}
is the effective Hamiltonian describing the atoms' self-evolution and
interaction with the laser field. Here $\delta=\omega_L-\omega_a$ is the laser
detuning from the atomic transition frequency, $\Omega_{1,2}$ are the complex
laser driving Rabi frequencies for each atom, $\hat \sigma^x_i,\,\hat
\sigma^y_i,\,\hat \sigma^z_i,\,\hat \sigma^\pm_i=\hat \sigma^x_i\pm i\hat
\sigma^y_i,\,i=1,2,$ are (using the well-known analogy between two-level atoms
and spins) the spin-$\frac{1}{2}$ Cartesian component and transition operators
of the $i$-th atom, and we define $g$, $f$ and $\gamma$ so that
$\gamma_{11}=\gamma_{22}=\gamma,\,\gamma_{12}=\gamma_{21}=g\gamma,\,\chi=f\gamma$.
The distance dependent parameters $g$ and $f$, describing, respectfully, the
photon exchange rate and coupling due to the RDDI, are defined differently for
different types of the atomic transition in question. Defining $p=0,\,q=2$ for
$\Delta m=0$ and $p=1,\,q=-1$ for $\Delta m=\pm 1$ transitions (with the
quantization axis coinciding with $\vec R$) and assuming that dipole matrix
elements of the atoms are collinear to each other and perpendicular to $\vec
R$, we have the following expressions for $g(\varphi)$ and $f(\varphi)$
\cite{milloni74}:
\begin{equation}\label{f&g}
  g(\varphi)= \frac{3}{2}\left(p\frac{\sin \varphi}{\varphi}+q\frac{\sin \varphi}{\varphi^3}-p\frac{\cos
  \varphi}{\varphi^2}\right),\quad
  f(\varphi)= \frac{3}{2}\left(p\frac{\cos \varphi}{\varphi}+q\frac{\cos \varphi}{\varphi^3}+q\frac{\sin
  \varphi}{\varphi^2}\right),
\end{equation}
where $\varphi=kR$ is the ``dimensionless distance'' between the atoms and
$k=\omega_a/c$. Throughout the rest of the article we will consider $\Delta
m=\pm 1$ case for determinacy as for the $\Delta m=0$ case the results are
qualitatively same.

It can easily be shown that the Dicke states, $\ket{\psi_e}=
\ket{e}_1\ket{e}_2,\, \ket{\psi_g}=\ket{g}_1\ket{g}_2,\,
\ket{\psi_s}=\frac{1}{\sqrt{2}}(\ket{g}_1\ket{e}_2+\ket{e}_1\ket{g}_2)$, and
$\ket{\psi_a}=\frac{1}{\sqrt{2}}(\ket{g}_1\ket{e}_2-\ket{e}_1\ket{g}_2)$ (where
$\ket{e}_i$ and $\ket{g}_i$ are the upper and lower levels of the $i$th atom)
are the eigenvectors of $\hat{\cal H}_{\rm eff}$ (when excluding the laser
driving), while the rest of the atomic dynamics can be described as radiative
decay to/from the antisymmetric $\ket{\psi_a}$ state with the rate
$\gamma_-=(1-g)\gamma$ and to/from the symmetric $\ket{\psi_s}$ one with the
rate $\gamma_+=(1+g)\gamma$.

Analytical stationary solutions of (\ref{ME}) can be found for the case of
symmetric excitation $\Omega_1=\Omega_2=\Omega$ (without loss of generality we
can assume that $\Omega$ here is real and positive) with the stationary
populations of the Dicke states given by

\begin{equation}\label{statpop}
\renewcommand{\arraystretch}{2.3}
\begin{array}{c}
   \displaystyle N_e=N_a=\frac{\Omega^4}{\left(\gamma^2+4\delta^2+2\Omega^2\right)^2 +
   \gamma(\gamma^2+4\delta^2)(f^2\gamma+g^2\gamma+2g\gamma-4f\delta)},\\
   \displaystyle N_s=\frac{2\Omega^2(2\gamma^2+8\delta^2+\Omega^2)}{\left(\gamma^2+4\delta^2+2\Omega^2\right)^2 +
   \gamma(\gamma^2+4\delta^2)(f^2\gamma+g^2\gamma+2g\gamma-4f\delta)},\\
   N_g=1-N_e-N_a-N_s.
\end{array}
\end{equation}
Graph $N_s(\Omega,\delta)$, corresponding to $\varphi=0.5$, is shown in Fig.
2a. The antisymmetric case, when the laser beam is parallel to $\vec R$, allows
no such simple analytical solution since in this case the relation between the
Rabi frequencies for two atoms is more complex
$\Omega_1=e^{i\varphi}\Omega_2=\Omega$ (although $\Omega$ here is again real
and positive). However, the numerical solution is easily obtained and the
corresponding dependence $N_a(\Omega,\delta)$ is shown in Fig. 2b.

If we aim to transfer the maximum amount of population into one of the
maximally entangled states $\ket{\Psi_a}$ or $\ket{\Psi_s}$ by a short coherent
pulse, a good criterion for finding optimal values of the laser field
parameters, $\Omega$ and $\delta$, is whether the population of the
corresponding level is close to 0.5 in the stationary solution. From the
analysis of (\ref{statpop}) and the graphs in Fig. 2 we deduce that the optimal
parameters can be well approximated by $\delta_{opt}=\pm\chi(\varphi)/2$ and
$\Omega_{opt}=\sqrt{|\chi(\varphi)|\gamma_\pm}$ with the upper/lower sign for
the symmetric/antisymmetric geometries, respectfully. We then have
$|\delta_{opt}|=|\chi|/2\gg\Omega_{opt}\gg\gamma_\pm$ for sufficiently small
distances, so that the transition of interest is saturated while at the same
time the Rabi frequency is moderate enough to avoid broadband excitation of the
Dicke level we are not interested in.

Given the optimal parameters obtained in the previous section we proceed to
find the fidelity of creation of the maximally entangled states, that is the
maximum amount of population one can transfer using {\em pulses} of radiation.
Considering the dynamics of the populations under optimal parameters laser
driving that is turned on at the time instant $t=0$ (when all of the population
is in $\ket{\Psi_g}$ state), we define the optimal pulse duration as the time
when the population of the state we are interested in, reaches its first
maximum. For the so chosen parameters we plot in Fig. 3a the populations
achieved by applying the optimal pulse, as a function of interatomic distance
$\varphi$. The populations approach unity for both geometries as $\varphi$ goes
to zero suggesting that almost perfect transfer of population is attainable at
small interatomic distances.

It is easy to explain this result. As $R$ goes to zero the energy splitting
between the symmetric and antisymmetric state, equal to $2\chi=2f\gamma$, grows
to infinity (of course, within the evident limitations of negligibility of
exchange effects \cite{kurizki87}). The ``parasitic'' excitation of the levels
we are not interested in is then avoided by shifting the laser frequency so
that we are in resonance only with one of the excited Dicke states. And if we
have in possession arbitrarily strong and arbitrarily tunable laser, as
$\varphi$ goes to zero we can produce shorter and shorter pulses thereby
decreasing the decay probability during the pulse.

In addition, in the case of the symmetric excitation laser driving matrix
element for the transitions involving the antisymmetric state vanishes. This
means that its population can only come from decay of the upper $\ket{\Psi_e}$
level. In fact, in the stationary solution (\ref{statpop}) the two states even
have the same populations, which are negligible for large detunings. In the
antisymmetric excitation case, however, the situation works against us. As the
interatomic distance goes to zero the Rabi frequencies of the two atoms become
closer in phase, diminishing along the way the matrix elements of transitions
involving antisymmetric state to zero. But even with such a small value of
excitation efficiency, we can still manage to transfer the population to the
antisymmetric state because the symmetric excitation remains far off resonance,
and  get as a reward the increased lifetime
$\tau=1/\left((1-g(R))\gamma\right)\gg 1/\gamma$ of our maximally entangled
state (using this ``durability'' of the antisymmetric $\ket{\Psi_a}$ state, we
can also create some entanglement passively as described in \cite{grishan98},
but it is difficult to obtain a high fidelity that way).

Now, instead of calculating different entanglement measures \cite{plenio98}
(and then figuring out which of them better suits our purposes), we will
consider how much our created states violate a simple Bell inequality. It can
be easily shown \cite{preskill} that for any classical local variable
distribution the probabilities of finding the atoms' ``spins'' aligned along
$\vec n_z$ direction after coherent rotations fulfill the following Bell
inequality:
\begin{equation}\label{bellineq}
  P_{\rm diff}(0,2\pi/3)+P_{\rm diff}(2\pi/3,-2\pi/3)+P_{\rm diff}(0,-2\pi/3)\geq 1,
\end{equation}
where the $P_{\rm diff}(\varphi_1,\varphi_2)$ is the probability of getting
different results of the measurements of the two spins, i.e., of finding one
spin aligned along the $\vec n_z$ direction and the second one against it after
the first spin is rotated by $\varphi_1$ and the second one by $\varphi_2$
around $OX$ or $OY$ axis. In quantum mechanics, however, the left hand side of
(\ref{bellineq}) amounts to only 0.75 for the pure maximally entangled
$\ket{\Psi_a}$ state. To apply the same Bell inequality (\ref{bellineq}) to the
case of the symmetric excitation geometry we first perform a $180^{\circ}$
rotation along $\vec n_z$-axis with one of the atoms (thus turning
$\ket{\Psi_s}$ state into $\ket{\Psi_a}$ one) and then perform the measurement
as described above. In Fig. 3b the l.h.s. of (\ref{bellineq}) is plotted
against the dimensionless distance $\varphi$ showing that for $\varphi$ less
than $\approx 0.5$ we have the violation of the inequality and as
$\varphi\rightarrow 0$ we recover the pure states limit of 0.75.

\medskip
In summary, while being admittedly unrealistic, our model offers a few insights
into how efficiently the RDDI can be used to entangle atoms or implement
quantum logic gates \cite{brennen98,deutsch95}. We have shown that considerable
fidelities (up to 0.8) of creation of one of the maximally entangled Dicke
states and Bell inequality violations can be realized if the atoms are placed
within distances of the order of a tenth of a wavelength of the working
transition. Such distances can be achieved, for example, in the ground
vibrational states of two atoms in optical lattices
\cite{meystre96,brennen98,wallis95}.

However, all of the practical applications (say, quantum teleportation
\cite{bennett93}) require stable entanglement.  Even the Bell inequalities
violation considered above can be verified only if the produced entanglement
lives sufficiently long so that the atoms can be spatially separated for
individual addressing and photodetection (imperfectness of the detectors
constitutes another problem that has not yet been addressed). Of course, this
cannot be achieved in our model, since the dipole interaction and the decay
have the same physical nature and we cannot  avoid the latter while making use
of the former.

Relatively stable coherences (with lifetimes of the order of seconds) can be
generated if we use the Zeeman sublevels of the atoms as the working levels
(qubits). The RDDI is negligible for them, but using three-level atoms in
$\Lambda$-configuration instead of two-level atoms we can overcome that
difficulty. We can use Raman pulses transferring the population between the
Zeeman sublevels via a higher lying ``transit'' Dicke level (STIRAP techniques
\cite{stiraprev} might be an alternative), so that the transitions between each
of the Zeeman sublevels states and the excited ``transit'' state benefit from
substantial RDDI. Then by choosing one-photon detunings to be in resonance with
only one of the higher level Dicke states we can generate the Zeeman sublevels
entanglement.

In this paper we present the first tentative quantitative model of the process
of entanglement of atoms with the help of the RDDI. Much remains to be done
before we can compare the results of the theory with possible practical
implementations, but the conclusions presented here are promising and therefore
encourage further theoretical developments.

This work was partially supported by Volkswagen Stiftung (grant No.\ 1/72944)
and the Russian Ministry of Science and Technical Policy.

\begin{figure}
\begin{center}
\epsfxsize=8cm \epsfclipon \leavevmode \epsffile{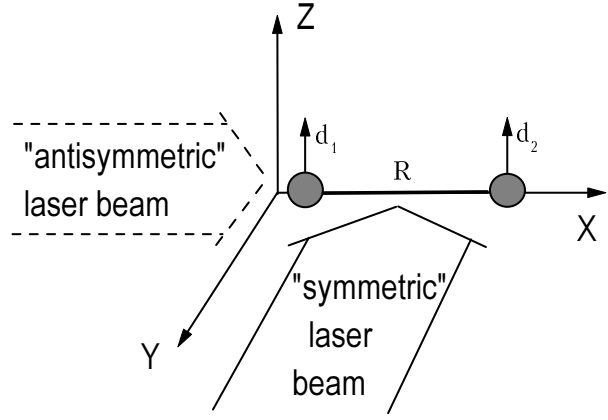}
\end{center}
\caption{Geometry of the model with shown directions of laser beams for the
symmetric and antisymmetric excitation.} \label{fig:fig1}
\end{figure}

\begin{figure}
\begin{center}
\epsfxsize=\textwidth \epsfclipon \leavevmode \epsffile{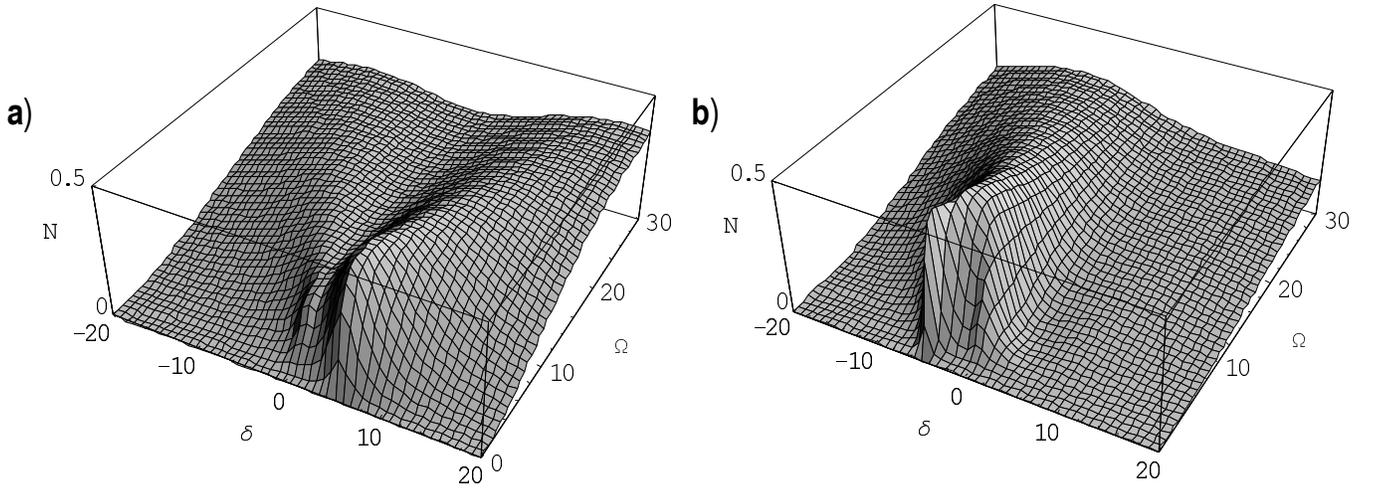}
\end{center}
\caption{The stationary populations of the symmetric (a) and  antisymmetric (b)
Dicke states versus the laser detuning $\delta$ and the laser driving Rabi
frequency $\Omega$ for  $\varphi= 0.5$. } \label{fig:fig2}
\end{figure}

\begin{figure}
\begin{center}
\epsfxsize=\textwidth \epsfclipon \leavevmode \epsffile{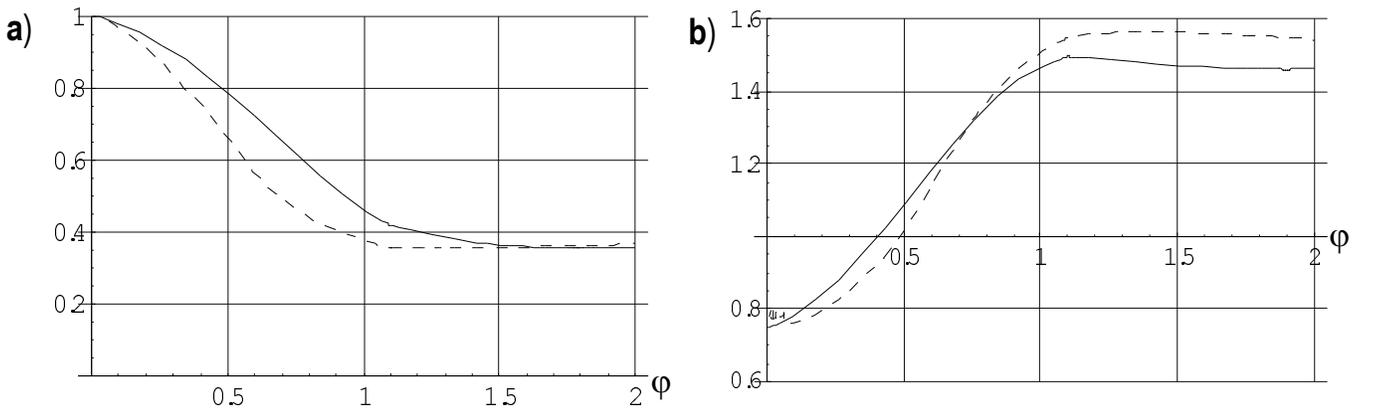}
\end{center}
\caption{a) The fidelity of creation of the maximally entangled states (by
applying the optimal pulse as described in text) versus interatomic distance
$\varphi$. b) The l.h.s. of the Bell inequality (5) versus $\varphi$. The pure
states limit is 0.75, uncorrelated states lead to 1.5, the classical boundary
is 1.0. Solid and dashed lines correspond to the antisymmetric and symmetric
cases, respectfully.} \label{fig:fig4}
\end{figure}


\begin{thebibliography}{99}
\vspace{-1.3cm}
\bibitem{preskill}
See, e. g., John Preskill, {\em Lecture notes on Physics 229: Quantum
information and computation}, located at Caltech web site \\
http://www.theory.caltech.edu/people/preskill/ph229/index.html.

\bibitem{atoment}
C. Cabrillo, J.I. Cirac, P. Garcia-Fernandez, and P. Zoller, Phys. Rev. {\bf A
59} (1999) 1025; M. B. Plenio, S. F. Huelga, A. Beige, P. L. Knight, Phys. Rev.
{\bf A 59} (1999) 2468; Q. A. Turchette, C. S. Wood, B. E. King, C. J. Myatt,
D. Leibfried, W. M. Itano, C. Monroe, D. J. Wineland, Phys. Rev. Lett. {\bf 81}
(1998) 3631; D. Jaksch, H.-J. Briegel, J.I. Cirac, C.W.Gardiner, and P. Zoller,
Phys. Rev. Lett. {\bf 82} (1999) 1975.

\bibitem{agarwal81}
G.S. Agarwal and L.M. Narducci, E. Apostolidis, Opt. Comm. {\bf 36} (1981) 285.

\bibitem{meystre93}
Georg Lenz and Pierre Meystre, Phys. Rev.{\bf A 48} (1993) 3365.

\bibitem{kurizki87}
Gershon Kurizki and Abraham Ben-Reuven, Phys. Rev. {\bf A 36} (1987) 90.


\bibitem{brewer95}
R.G. Brewer, Phys. Rev. {\bf A 52} (1995) 2965.

\bibitem{andreevbook}
A.V. Andreev, V.I. Emel'yanov, Yu. A. Il'inskii, {\em Cooperative effects in
optics: superradiance and phase transitions}, Philadelphia, PA: Institute of
Physics Publishing, 1993.

\bibitem{beige98}
Almut Beige and Gerhard C. Hegerfeldt, Phys. Rev. {\bf A 58} (1998) 4133.

\bibitem{smith91}
A.M. Smith and K. Burnett, J. Opt. Soc. Am. { \bf B 8} (1991) 1592.

\bibitem{meystre96}
E.V. Goldstein, P. Pax and P. Meystre, Phys. Rev. {\bf A 53} (1996) 2604.

\bibitem{barenco95}
Adriano Barenco, David Deutsch, Arthur Ekert, Richard Jozsa, Phys. Rev. Lett.
{\bf 74} (1995) 4083.

\bibitem{brennen98}
Gavin K. Brennen, Carlton M. Caves, Poul S. Jessen and Ivan H. Deutsch, Phys.
Rev. Lett. {\bf 82} (1999) 1060.

\bibitem{milloni74}
P.W. Milloni and P. L. Knight, Phys. Rev. {\bf A 10} (1974) 1096.

\bibitem{grishan98}
B.A. Grishanin, V.N. Zadkov, Laser Physics {\bf 8} (1998) 1074.

\bibitem{plenio98}
Jens Eisert, Martin B. Plenio, quant-ph/9807034.

\bibitem{deutsch95}
D. Deutsch, A. Barenco, A. Ekert, Proc. Roy. Soc. of London {\bf A 449} (1995)
669.

\bibitem{wallis95}
H. Wallis, Phys. Rep. {\bf 255} (1995) 203.

\bibitem{bennett93}
Charles H. Bennett, Gilles Brassard, Claude Crepeau, Richard Jozsa, Asher
Peres, and Willam K. Wooters, Phys. Rev. Lett. {\bf 70} (1993) 1895.

\bibitem{stiraprev}
K. Bergmann, H. Theuer, and B. W. Shore, Rev. Mod. Phys. {\bf 70} (1998) 1003.



\end{thebibliography}
\end{document}